%% file: main.tex
\documentclass[letterpaper,twocolumn,10pt]{article}
\usepackage{usenix2019_v3}

\usepackage{tikz}
\usepackage{amsmath}

\usepackage{filecontents}

\usepackage{enumitem}
\usepackage{comment}
\usepackage{microtype}
\usepackage{graphicx}
\usepackage{subfigure}
\usepackage{booktabs} 

\usepackage{hyperref}
\usepackage{bm}

\usepackage{amsmath}
\usepackage{amssymb}
\usepackage{mathtools}
\usepackage{amsthm}

\usepackage{multirow}
\usepackage{algorithm}
\usepackage{algorithmic}
\usepackage{titling}

\usepackage[capitalize,noabbrev]{cleveref}

\theoremstyle{plain}

\theoremstyle{definition}

\theoremstyle{remark}

\usepackage[textsize=tiny]{todonotes}

\usepackage{xspace}

\newcommand{\red}[1]{\textcolor{black}{#1}}
\newcommand{\proj}{GNNPipe\xspace}
\newcommand{\avgspeedupgraphparallel}[0]{$1.58\times$\xspace}
\newcommand{\maxspeedupgraphparallel}[0]{$2.45\times$\xspace}
\newcommand{\maxspeedupdgl}[0]{$61.0\times$\xspace}
\newcommand{\avgcommvolumereduction}[0]{$8.69\times$\xspace}
\newcommand{\maxcommvolumereduction}[0]{$22.89\times$\xspace}
\newcommand{\avgcommoverheadreduction}[0]{$11.60\times$\xspace}
\newcommand{\maxcommoverheadreduction}[0]{$27.21\times$\xspace}

\begin{document}

\date{}

\title{\Large \bf \proj: Scaling Deep GNN Training with Pipelined Model Parallelism}

\author{
{\rm Jingji Chen}\\
Purdue University
\and
{\rm Zhuoming Chen}\\
Carnegie Mellon University
\and
{\rm Xuehai Qian}\\
Purdue University
} 

\maketitle

\begin{abstract}

Communication is a key bottleneck for distributed
graph neural network (GNN) training. 
This paper proposes {\em \proj}, a new approach
that scales the distributed 
full-graph {\em deep} GNN training. 
Being the first to use layer-level 
model parallelism for GNN training, \proj 
partitions GNN layers among GPUs, each device performs the computation
for a disjoint subset of consecutive GNN layers
on the whole graph. Compared to graph 
parallelism with each GPU handling a graph 
partition, \proj reduces the communication
volume by a factor of the number of GNN layers. 
\proj overcomes the unique challenges for 
pipelined layer-level model parallelism 
on the whole graph by partitioning it into dependent chunks, allowing the
use of historical vertex embeddings, and
applying specific training techniques to ensure convergence. We also propose a hybrid
approach by combining \proj with graph parallelism to handle large graphs, achieve 
better computer resource utilization and 
ensure model convergence. 
We build a general GNN training
system supporting all three parallelism setting.
Extensive experiments show that our method reduces the per-epoch training time by up to \maxspeedupgraphparallel (on average \avgspeedupgraphparallel) 
and reduces the communication volume and overhead by up to \maxcommvolumereduction and \maxcommoverheadreduction (on average \avgcommvolumereduction and \avgcommoverheadreduction),
respectively,
while achieving a comparable level of model accuracy and convergence speed compared to graph parallelism.

\end{abstract}

\input{intro.tex}
\input{background.tex}

\input{methods.tex}

\input{experiments.tex}
\input{related_works.tex}
\input{conclusion.tex}

\bibliographystyle{plain}
\bibliography{gnn}

\end{document}

%% file: intro.tex
\section{Introduction}

The past few years have witnessed the great success of graph neural networks (GNN), one of the fastest-growing subareas in neural network research community~\cite{hamilton2020graph}, in learning relational information from non-euclidean graph-structure data for various tasks
by combining message passing with traditional neural network layers~\cite{kipf2016semi,hamilton2017inductive,velivckovic2017graph,xu2018powerful,li2019deepgcns,chiang2019cluster,zeng2019graphsaint,chen2020simple}.
Inspired by
the success of the deep neural network models like convolution neural 
networks~\cite{he2016deep,simonyan2014very,szegedy2015going,szegedy2016rethinking,huang2017densely} in deep learning, tremendous research efforts have been devoted to {\em making GNN models deeper}~\cite{li2019deepgcns,chiang2019cluster,zhou2020towards,li2021gnn1000,chen2020simple,xu2018representation,rong2019dropedge,hasanzadeh2020bayesian,jia2020improving}. Based on recent studies, with sophisticated model architectural designs 
, deep GNN models with far more than two layers, e.g., {\em 64 layers}, are able to achieve new state-of-the-art results on various tasks~\cite{li2021gnn1000,jia2020improving,chen2020simple,li2021deepgcns,li2019deepgcns}, such as point cloud segmentation~\cite{wang2019dynamic}, graph inductive learning~\cite{hamilton2017inductive}, and node 
classification~\cite{kipf2016semi}. 
In particular, deep GNN models are required for tasks that require large 
receptive fields. An important example is neural subgraph matching~\cite{lou2020neural}, which discovers whether a subgraph
structure exists in a large graph.
In this application, the number of GNN layers should be at least
the diameter of the subgraph to ensure a sufficiently large receptive field~\cite{lou2020neural}. 

To realize the potential of deep GNNs, it is crucial to {\em scale deep GNN
training} since it demands 
prohibitively large amount of memory and computation resources that are beyond the capacity of a single GPU.
Unfortunately, 
{\em no existing GNN training methods can scale with and efficiently 
support deep GNN training}, which is the key motivation of this work. 
The goal of the paper is to propose {\em the first high performance
deep GNN training system}.

There are currently two GNN training methods. 
The {\em sampling-based} methods aim to reduce the amount of computation and memory resources required for training by using graph samples as the training
data.
Instead of processing the full graph data,
in each training iteration, the
sampling-based methods randomly sample a number of subgraphs 
that fits into a single GPU~\cite{hamilton2017inductive,chen2017stochastic,chen2018fastgcn,zou2019layer,chiang2019cluster,zeng2019graphsaint,zeng2021decoupling}.
In general, for a $\mathcal{L}$-layer GNN, each subgraph contains the sampled 
$\mathcal{L}$-hop subgraph from a vertex, in which each hop chooses a subset of neighbors.
The sampling-based methods can be naturally implemented with mini-batch
approach, where each batch contains a number of sampled subgraphs that 
can fit into the GPU memory.

The mini-batch approach suffers from two inherent drawbacks,
one of them is particularly relevant to deep GNNs. 
With multiple GPUs, the graph can be partitioned among them. 
Sampling the k-hop graph will incur inter-GPU communications 
when a sampled vertex resides in a remote GPU.
First, the method introduces {\em redundant computation and communication}.
Intuitively, it is caused by the fact that the two $\mathcal{L}$-hop subgraphs
from two vertices $v1$ and $v2$ can overlap, and the overlapped vertices and 
edges increase quickly with $\mathcal{L}$. 
We will explain the problem with a concrete example in \red{Section~\ref{sec:mini_batch}}.
Second, for a $\mathcal{L}$-layer deep GNN, the $\mathcal{L}$-hop subgraph from a vertex $v$ can be 
prohibitively large, e.g., \red{a 7-hop subgraph in Flickr~\cite{zeng2019graphsaint} can 
contain all vertices}.
With fixed memory budget for each subgraph, the sample rate needs to be
exceedingly small, otherwise, 
the subgraph size will be large and may not fit into the GPU memory.
It makes the subgraph fail to capture the property of the
original graph. 
While the mini-batch approach is used in several popular GNN training 
systems such as DGL~\cite{wang2019deep}, P3~\cite{gandhi2021p3}, \red{Legion~\cite{sun2023legion}, and many other recent systems~\cite{zheng2022bytegnn,liu2023bgl,polisetty2023gsplit,yang2022gnnlab,zhang2023ducati}
}, 
these systems can only support shallow GNNs with a small number of layers. 

The second GNN training method is the {\em full-graph} training,
which iteratively processes the entire input graph. 
This approach resembles the traditional graph processing, which propagates the information in the graph incrementally between directly connected vertices.
For GNNs, the property of a vertex is its {\em embedding},
the forward process of an {\em epoch} processes the entire graph data on the $\mathcal{L}$-layer GNN through {$\mathcal{L}$ iterations}, propagating the 
embeddings to $\mathcal{L}$-hop neighbors of each vertex. 
The backward process is similar. 
For a deep GNN, the number of layers determines the number of iterations,
however, the full-graph training
fails to scale with the number of GPUs due to the {\em severe communication bottleneck}.

With multiple GPUs, the full-graph training can be parallelized with 
{\em graph parallelism}~\cite{ma2019neugraph,wan2022pipegcn,thorpe2021dorylus,jia2020improving,wan2022bns,mostafa2022sequential} to distribute the workload across GPUs:
the full graph is partitioned into multiple subgraphs,
each GPU keeps one partition and the whole model (i.e., weights). 
The GPUs collaboratively 
train the GNN by training the model on the local graph partition.
The performance of graph parallel execution is heavily affected
by the amount of communication between GPUs, which is a function of 
graph partition. 
The graph communication is the main source of communication caused by the \textit{neural message passing}~\cite{gilmer2017neural}
for \textit{each} layer
to pass message---the embedding vectors of remote vertice---across the boundaries
among the subgraphs in different GPUs.
In \red{Section~\ref{sec:back_dist_training}}, we demonstrate that the worst-case 
aggregated communication volume to train a $\mathcal{L}$-layer GNN with
$\mathcal{H}$ hidden units using 
$\mathcal{M}$ GPUs on a graph with $\mathcal{N}$ vertices is
$O(\mathcal{LMNH})$.
Our experiments confirms the high communication overhead of distributed GNN 
training with graph parallelism. 
Table~\ref{tab:comm_overhead_moti} indicates that
the communication can take up to 86.26\% of the total training time.

\begin{figure}[htbp]
    \centering
    \includegraphics[width=0.8\linewidth]{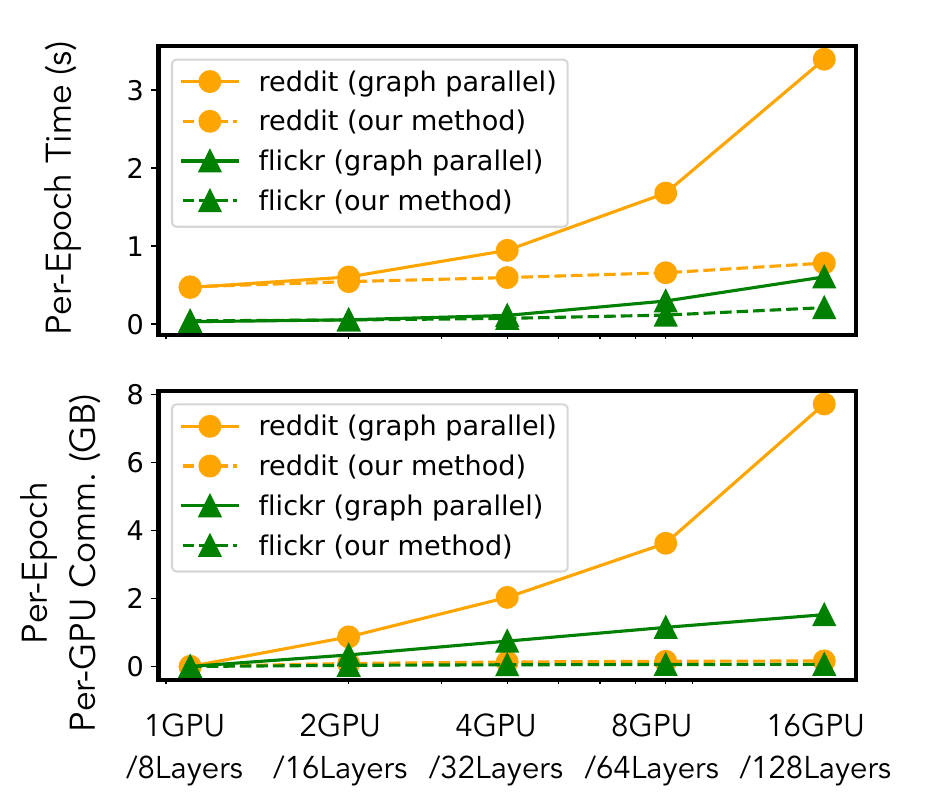}
    \caption{GNN training with graph parallelism exhibits poor scalability. When increasing both the number of GPUs and the workload size (number of layers) concurrently, the training time gets slower and the per-GPU communication volume gets higher (model: GCN, datasets: Reddit~\cite{hamilton2017inductive}, Flickr~\cite{zeng2019graphsaint}).}
    \label{fig:weak_scaling}
\end{figure}

The analysis results explain why the graph parallel
GNN training is {\em fundamentally not scalable}, particularly for deep GNNs.
In general, multiple GPUs increases not only the aggregated computation
capability but also the aggregated communication bandwidth, \red{roughly 
linear with the number of GPUs}.
If the computation and communication amount of an application increases (almost)
linearly with the number of GPUs, 
then the application is scalable. 
However, the above communication complexity indicates that the 
communication volume for GNN training 
increases with {\em both} the number of GPUs and 
the number of GNN layers.
\red{Figure~\ref{fig:weak_scaling} experimentally confirms our analysis
by showing the performance of graph parallel GNN training 
with increase number of layers}.

To enable efficient distributed deep GNN training, we propose {\em \proj}, 
a new full-graph training
method based on {\em layer-level model parallelism}~\cite{krizhevsky2014one}.
\proj partitions GNN layers among GPUs, each device
is responsible for performing the computation for 
{\em a disjoint subset of consecutive GNN layers} on the {\em whole} graph.
During training, each of the $\mathcal{M}$ GPUs only executes {\em one} iteration
and communicates with {\em one} remote GPU with the updated embeddings
of all vertices.
Thus, the total communication volume increase linearly with 
$\mathcal{M}$, and the communication complexity is $O(\mathcal{MNH})$,
reducing $O(\mathcal{LMNH})$ of graph parallelism by a factor of $\mathcal{L}$.

Using pipeline in machine learning model training 
is not a new idea, 
why has this idea not been adopted for GNN training?
The fundamental challenge is: 
each iteration on the whole graph 
needs to be executed {\em sequentially} according
to GNN's layer structure by different GPUs.
Thus, at any point in time, only {\em one} GPU is utilized.
In non-GNN model training, the training samples
in a batch are {\em independent}, thus, a batch can be divided into multiple
independent ``micro-batches''~\cite{huang2019gpipe}. The pipeline stages
can achieve parallelism for the batch by processing different micro-batches
concurrently. 
For GNN training, vertices in the graph is connected and they are 
{\em dependent}. 

To tackle the challenge, we propose to partition the whole graph
into {\em chunks} and allow the {\em stale} historical embedding during
\proj's pipelined layer-level model parallelism training.
The chunks are similar to micro-batches in GPipe~\cite{huang2019gpipe} except
that they are {\em dependent}.
When we achieve parallelism among chunks, depending on the graph partition 
and the order of processing all chunks, some may contain the stale
historical vertex embedding, which may affect the training efficiency or
even lead to non-convergence.
To ``recover'' the training efficiency when pipelining dependent chunks,
we propose three techniques: (1) chunk shuffling to avoid systematical
bias; (2) fixing historical embeddings to ensure training stability;
(3) avoid using historical gradients which tend to incur large errors.
With the three training techniques, \proj is able to successful
train deep GNN models 
\red{without accuracy loss.}
Fundamentally, our solution co-designs training algorithm with 
system and slightly
trades statistical efficiency for execution efficiency. 

As a step further, 
we develop a general {\em hybrid} training method by 
combining layer-level model parallelism and graph parallelism, in which 
they can be considered as special cases. 
Specifically, each GPU can be assigned to process a subset of consecutive
layers on a graph partition, instead of the whole graph. 
The hybrid parallelism addresses three practical issues:
(1) the large graphs may not fit into a single GPU's memory;
(2) the number of layers can be less than the 
number of available GPUs; and 
(3) for some deep GNNs, using deep pipeline may affect convergence.
\red{Section~\ref{sec:hybrid_analysis}} also analyzes the trade-offs of communication volume
with different parallelism settings. 

To the best of our knowledge, 
\proj is the {\em first} GNN training method 
that exploits layer-level model parallelism 
and achieves
performance superiority over solutions based on graph parallelism.
The hybrid approach enables a general GNN training system 
that can efficiently work with both shallow and deep GNNs,
explore the trade-off between training and execution efficiency, and 
utilize all available GPU computing resources.

One can argue that, the advanced NVLink~\cite{nvlink} with \red{up to 900 GBps} 
bandwidth can largely 
mitigate such communication bottleneck. 
However, we must at the same time consider {\em high cost} of the advanced 
machines with NV-Link. 
For example, NVIDIA's DGX A100 server with 8 NVLink-connected A100 GPUs costs 200K. 
Thus, it is important to support deep GNN training (and machine learning in
general) in a {\em cost-effective}
manner with affordable solutions whenever possible.
This paper is an important step toward such goal.

\begin{table}[t]
    \centering
    \caption{Communication overhead (the ratio between the communication time and the overall training time) of our graph-parallel baseline. Evaluated on the Reddit dataset~\cite{hamilton2017inductive} with NVIDIA A100 GPUs interconnected by a 100Gbps InfiniBand network. The GNN model is a 3-layer GCN with 256 hidden units.}
    \begin{tabular}{c|ccc}
        \hline
         Num.GPU & 4 & 8 & 12 \\
         \hline
         Comm.Time/Runtime & {69.13\%} & {79.29\%} & {86.26\%} \\
         \hline
    \end{tabular}
    \label{tab:comm_overhead_moti}
\end{table}

\red{
We extensively evaluated \proj
with four 32-layer GNN models
on a 8-GPU cluster with a fast
200Gbps InfiniBand network.
The experiments show that,
comparing with our graph-parallel baseline,
\proj significantly reduces the communication volume and overhead to up to \maxcommvolumereduction  and \maxcommoverheadreduction (on average \avgcommvolumereduction and \avgcommoverheadreduction),
and improves the per-epoch training time by up to \maxspeedupgraphparallel (on average \avgspeedupgraphparallel).
It also outperforms DGL~\cite{wang2019deep},
a state-of-the-art mini-batch-based distributed system,
by up to \maxspeedupdgl.
}

%% file: background.tex
\section{Background}

\subsection{Graph Neural Networks Basics}

Given a graph $\mathcal{G=(V,E)}$,
the goal of an $\mathcal{L}$-layer graph neural network (GNN) is to learn an 
embedding vector representation $\bm{h}_v = \bm{h}_v^{(\mathcal{L})}$ for each $v\in \mathcal{V}$.
The embedding $\bm{h}_v^{(\ell)}$ at layer $\ell$ can be obtained
by $\bm{h}_v^{(\ell-1)}$ with the differentiable $\textsc{AGGREGATE}^{(\ell)}(\cdot)$ and $\textsc{UPDATE}^{(\ell)}(\cdot)$ functions.
We describe the process mathematically as follows.
\begin{equation}
    \begin{split}
    \bm{z}_v^{(\ell)} &= \textsc{AGGREGATE}^{(\ell)}(\{\bm{h}_u^{(\ell-1)}|u\in \mathcal{N}(v)\}) \\
    \bm{h}_v^{(\ell)} &= \textsc{UPDATE}^{(\ell)}(\bm{h}_v^{(\ell-1)}, \bm{z}_v^{(\ell)}) 
    \end{split}
\end{equation}
where $\mathcal{N}(v)$ contains the neighbor vertices of $v$ and 
the input to the first layer $\bm{h}_v^{(0)}$ equals the feature vector of vertex $v$.
As an example, 
for the GCN model~\cite{kipf2016semi},
the aggregation operation is a weighted sum of the neighbor embeddings,
which is 
$\textsc{AGGREGATE}^{(\ell)} = \sum_{u\in \mathcal{N}(v)}\frac{1}{\sqrt{\mathcal{D}_v\mathcal{D}_u}}\bm{h}_u^{(\ell-1)}$ ($\mathcal{D}_v$ and $\mathcal{D}_u$ is the degree of $v$ and $u$),
while $\textsc{UPDATE}^{(\ell)}(\cdot)$ is a fully-connected neural network layer.


\subsection{Distributed Mini-batch GNN Training}
\label{sec:mini_batch}


Mini-batch based distributed GNN training 
is inherited from traditional machine learning systems.
This method first divides the vertices $\mathcal{V}$
into a large number of fine-grained vertex sets called mini-batches (denoted as $\mathcal{M}_1$, $\mathcal{M}_2$, $\ldots$).
During each training iteration,
each GPU will be responsible for generating the final-layer embeddings of a different mini-batch $\mathcal{M}_i$ concurrently (i.e., calculating all $\bm{h}_v^{(\mathcal{L})}$ s.t. $v\in \mathcal{M}_i$). 
Recall that $\bm{h}_v^{(\ell)}$ depends on 
all $\bm{h}^{(\ell-1)}$ of $v$'s one-hop neighbors,
which, similarly, depend on the $\bm{h}^{(\ell-2)}$ of all $v$'s two-hop neighbors.
Due to the recursive dependency,
to calculate $\bm{h}_v^{(\mathcal{L})}$ s.t. $v\in \mathcal{M}_i$,
the GPU needs to load the vertex data (i.e., $\bm{h}^{(0)}$) of the $\mathcal{L}$-hop subgraph of $v$,
and then calculate $\bm{h}^{(1)}$, $\bm{h}^{(2)}$,
$\ldots$,
$\bm{h}^{(\mathcal{L})}$ sequentially.
For an example,
assume that 2 GPUs are used to process the graph shown in 
Figure~\ref{fig:dist_training_example}, 
and there are two GNN layers ($\mathcal{L}=2$).
GPU 1 is responsible for calculating mini-batch $\mathcal{M}_1=\{v_0, v_2\}$ while GPU 2 processes mini-batch $\mathcal{M}_2=\{v_1,v_3\}$.
GPU 1 will need to load the 2-hop subgraph of $v_0$ and $v_2$, which is $\{v_0, v_1, v_2, v_3, v_4\}$,
while GPU 2 will load the 2-hop subgraph of $v_1$ and $v_3$ that consists of $\{v_0,v_1,v_2,v_3,v_4,v_5,v_6\}$.
Both GPUs will use the loaded data to calculate their own mini-batches independently.
It is important to note that the $\mathcal{L}$-hop subgraphs of the mini-batches processed by different GPUs are likely to overlap (e.g., the $\mathcal{L}$-hop subgraph of $\mathcal{M}_1$ and $\mathcal{M}_2$ have five overlapped vertices),
which fundamentally causes redundant data loading and computation in the mini-batch based training method.


\subsection{Distributed Full-Graph GNN Training with Graph Parallelism}
\label{sec:back_dist_training}

A more natural and efficient way to implement 
full-graph training is to explore graph parallelism. 
In this approach, the graph is partitioned and stored in the memory 
of each GPU, which is only responsible for processing the local partition. 
It provides a way to leverage the increasing amount of 
memory and compute resource.

\begin{figure}
    \centering
    \includegraphics[width=.9\linewidth]{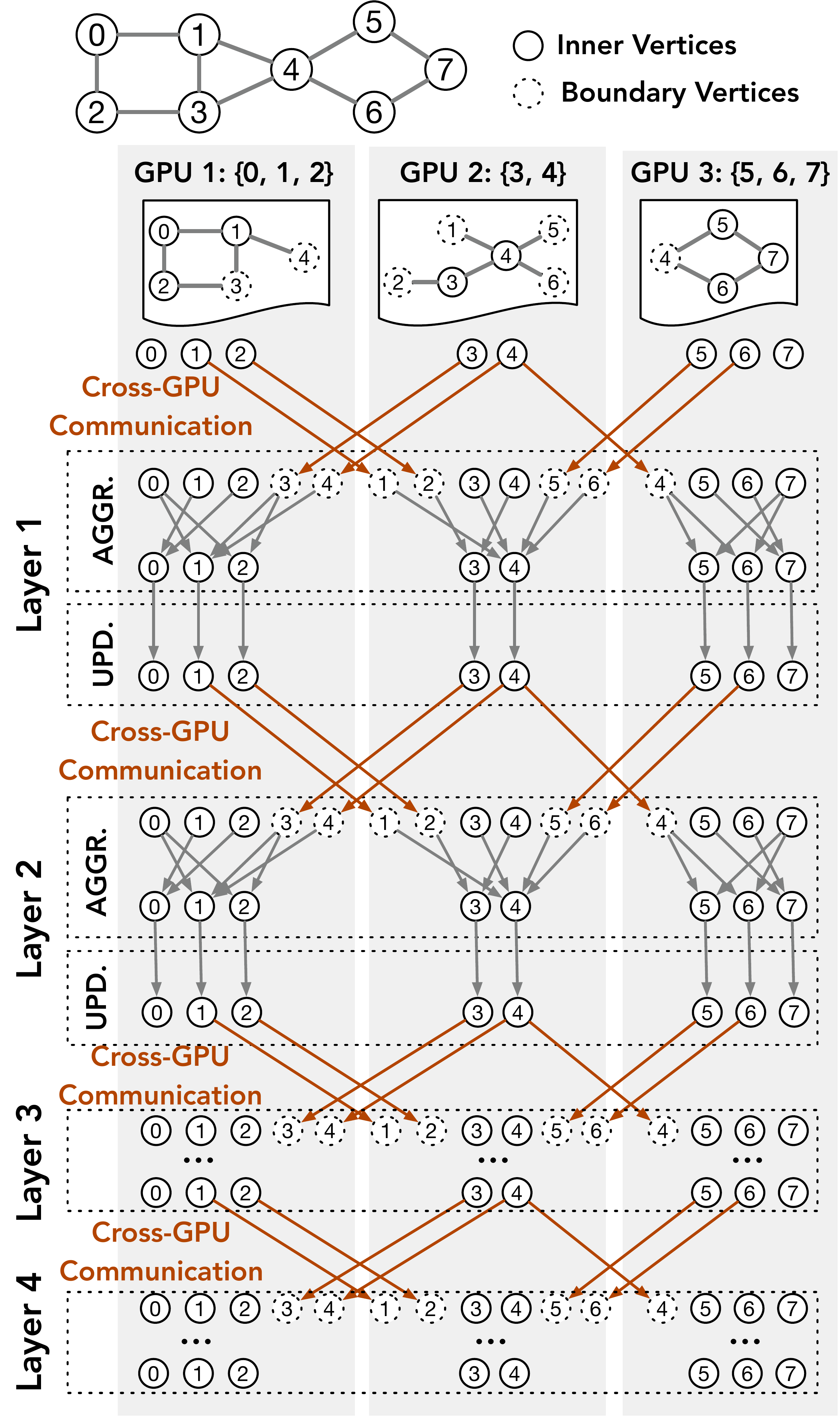}
    \caption{Distributed full-graph GNN training with graph parallelism (the backward pass omitted for simplicity).}
    \label{fig:dist_training_example}
\end{figure}

With $\mathcal{M}$ GPUs, the set of all vertices $\mathcal{V}$ is divided into $M$ non-overlapping vertices partitions $\mathcal{V}_1$, $\mathcal{V}_2$, $\ldots$, $\mathcal{V}_{\mathcal{M}}$ such that $\bigcup_{i=1}^{\mathcal{M}} \mathcal{V}_{i} = \mathcal{V}$.
$\mathcal{V}_i$ is called the \textit{inner vertices}~\cite{shao2022distributed} of GPU $i$.
Each GPU is responsible for calculating the embeddings and corresponding gradients of its own inner vertices.
We define the \textit{boundary vertices}~\cite{shao2022distributed} of GPU $i$ as $\mathcal{B}_i = \bigcup_{v\in \mathcal{V}_i}\mathcal{N}(v)-\mathcal{V}_i$,
which are the vertices outside $\mathcal{V}_i$ that are $\mathcal{V}_i$'s directed neighbors.
The embeddings of boundary vertices $\mathcal{B}_i$ are necessary to calculate the embeddings of $\mathcal{V}_i$.
Since $\mathcal{B}_i$'s embeddings are produced by a GPU other than $i$,
they have to be moved from the remote producer GPU to GPU $i$,
which incurs \textit{cross-GPU communication}.

Figure~\ref{fig:dist_training_example} illustrates the process using an
example with three GPUs and a 4-layer GNN. 
The full graph is divided into three partitions 
$\mathcal{V}_1=\{v_0, v_1, v_2\}$, $\mathcal{V}_2 = \{v_3, v_4\}$ and $\mathcal{V}_3 = \{v_5, v_6, v_7\}$.
The boundary vertices of each GPU are $\mathcal{B}_1 = \{v_3, v_4\}$, $\mathcal{B}_2 = \{v_1, v_2, v_5, v_6\}$, $\mathcal{B}_3 = \{v_4\}$, respectively.
At the beginning of each layer $\ell$, 
GPU $i$ needs to retrieve the previous-layer embeddings
of $\mathcal{B}_i$ from other GPUs (i.e., GPU $3$ fetches the embeddings $\bm{h}_4^{(\ell - 1)}$ from GPU $2$).
Once the cross-GPU communication is completed, 
each GPU can start calculating $\bm{h}_v^{(\ell)}$ locally with $\textsc{AGGREGATE}^{\ell}(\cdot)$ and $\textsc{UPDATE}^{\ell}(\cdot)$.

Unlike the mini-batch approach, the large number of GNN layers do not make
graph parallelism obviously infeasible, 
since each layer corresponds to an iteration
similar to the one in graph processing. 
However, it is the exceedingly high communication cost that makes graph
parallelism fundamentally not scalable for deep GNNs. 
We will explain the analysis in the next section before introducing the 
layer-level model parallelism.

%% file: methods.tex
\section{\proj Approach}
\label{sec:method}

\subsection{Motivation}



The key motivation of this work is the high communication complexity 
of graph parallelism. 
According to the discussion earlier, 
at the beginning of each layer $\ell$,
GPU $i$ needs to fetch all $\bm{h}_v^{(\ell - 1)}$ for $v\in \mathcal{B}_i$ from other GPUs.
Hence, the total communication cost per layer is $\mathcal{H}\sum_{i=1}^{\mathcal{M}} |\mathcal{B}_i|$ floating points where $\mathcal{H}$ is the hidden dimension. 
For simplicity, we assumed that the hidden dimensions across all layers are the same.
The total communication volume of a forward pass is thus $\mathcal{LH}\sum_{i=1}^{\mathcal{M}} |\mathcal{B}_i|$ for an $\mathcal{L}$-layer GNN. 
The communication cost of 
the backward pass is the same.
The worst case is reached when each vertex outside $\mathcal{V}_i$ is adjacent to at least one vertex in $\mathcal{V}_i$, i.e., 
$\mathcal{B}_i = \mathcal{V} - \mathcal{V}_i$, and 
the communication volume can be as high as $O(\mathcal{LH}\sum_{i=1}^{\mathcal{M}}(|\mathcal{V}|-|\mathcal{V}_i|)) = O(\mathcal{LH|V|}(\mathcal{M}-1)) = O(\mathcal{LHNM})$ assuming $\mathcal{|V|}=\mathcal{N}$.
It means that,
\textit{at each layer},
\textit{all embeddings} produced by a GPU needs to be sent to \textit{all other GPUs},
introducing exceedingly high communication overhead.

Most importantly, while the above communication complexity 
analysis is based on worst-case scenario, it is not difficult to achieve that 
in practical setting. 
In the following, we explain that, 
even with a \textit{perfect} partitioner,
distributed training on a \textit{highly sparse} graph
can incur communication cost close to the worst-case complexity.
We define the perfect partitioner as one that can partition $\mathcal{V}$ into $\mathcal{M}$ partitions $\mathcal{V}_1$, $\mathcal{V}_2$, $\ldots$, $\mathcal{V}_{\mathcal{M}}$ with a minimum communication cost while ensuring load balancing (i.e., $|\mathcal{V}_i|\approx \mathcal{N}/\mathcal{M}$ for each $i$).
Consider a random sparse graph model $\mathcal{G}=(\mathcal{V}, 
\mathcal{E})$ with $\mathcal{N}$ vertices.
We assume that, given any pair of vertices $u,v\in \mathcal{V}$, 
they are directly connected by a probability $p$ independently.
Hence, given a vertex $w\notin \mathcal{V}_i$,
the probability that it is directly connected to at least
one vertex in $\mathcal{V}_i$ is $1-(1-p)^{|\mathcal{V}_i|}$.
As a result, the expected number of boundary vertices in GPU $i$ is $E[|\mathcal{B}_i|] = (|\mathcal{V}|-|\mathcal{V}_i|)(1-(1-p)^{|\mathcal{V}_i|})\approx (|\mathcal{V}| - |\mathcal{V}_i|)(1-(1-p)^{\mathcal{N}/\mathcal{M}})$.
With $\mathcal{N}=10^6$, $\mathcal{M}=8$, and $p=2\times 10^{-5}$ (the average degree is $20$),
$E[|\mathcal{B}_i|] \approx 0.92(|\mathcal{V}| - |\mathcal{V}_i|)$,
which is very close to $|\mathcal{V}| - |\mathcal{V}_i|)$
in the worst-case.


\subsection{Solution: Layer-Level Model Parallelism}
\label{sec:pipeline_parallelism}

\begin{figure}
    \centering
    \includegraphics[width=\linewidth]{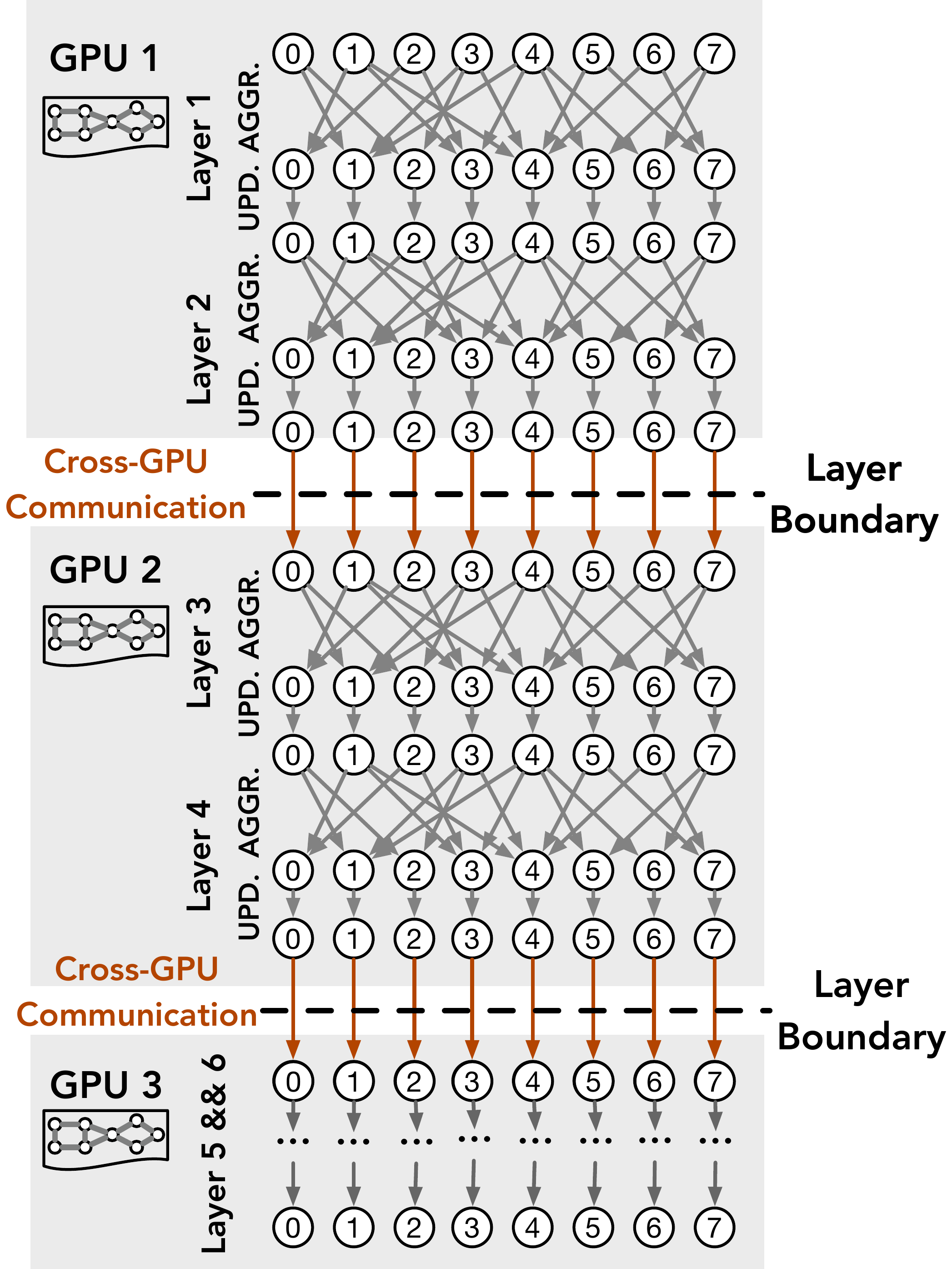}
    \caption{Distributed training a 6-layer GNN with model parallelism (the backward pass is omitted for simplicity). The graph is the same as Figure~\ref{fig:dist_training_example}.}
    \label{fig:dist_training_model_parallel}
\end{figure}

To avoid the large communication overhead in graph parallelism, 
we propose to exploit layer-level model parallelism~\cite{krizhevsky2014one}
for distributed GNN training. 
Unlike graph parallelism, the model parallelism 
partitions layers rather than the graph, and 
each GPU is responsible to train a subset of consecutive layers.
Figure~\ref{fig:dist_training_model_parallel} illustrates the 
idea with an example. 
The 6-layer GNN is partitioned and distributed among three GPUs; and
GPU $1$, GPU $2$ and GPU $3$ train layers 1-2, 3-4, and 5-6, respectively. 
In this organization, the inter-GPU communication occurs at the layer boundaries:
since layer 3 takes the embeddings of all vertices produced by layer 2 as its input,
these embeddings will be transferred from GPU $1$ to GPU $2$
with cross-GPU communication.
For deep GNNs~\cite{chen2020simple,li2019deepgcns} with large number of layers, 
the layer-level model parallelism can naturally expose abundant 
parallelism in the layer dimension.


This new approach provides two major advantages:
{\em less} communication volume and 
more {\em balanced and predictable} communication pattern.
In layer-level model parallelism, since the 
communication occurs at layer boundaries, 
only embeddings produced by a boundary layer, e.g., layer-2/4 embeddings in Figure~\ref{fig:dist_training_model_parallel},
need to be transferred, only to {\em one} remote GPU.
The data sent across each layer boundary 
is $\mathcal{NH}$ floating point values.
Since there are $\mathcal{M}-1$ layer boundaries,
the total communication volume for layer-level model parallelism
is $O((\mathcal{M}-1)\mathcal{NH}) = O(\mathcal{MNH})$,
which reduces that of graph parallelism ($O(\mathcal{LHNM})$) by a factor of $O(\mathcal{L})$.
Not only that the communication volume is lower for model parallelism, but 
also the communication pattern is more balanced and predictable, which 
makes the above complexity precisely capture the worst-case.
In comparison, for graph parallelism,
in the worst case,
all embeddings produced by each layer need to be sent to {\em all} other GPUs.

The idea of model and pipeline parallelism has been recently proposed
and extensively studied in distributed 
machine learning model training~\cite{narayanan2019pipedream,huang2019gpipe}.
However, we are not simply applying the idea to yet another scenario since 
the nature of GNN training inherently brings a unique challenge
despite its low communication cost.
The key difficulty is the {\em sequential} execution nature 
of full-graph GNN training.
Specifically, the embeddings need to be calculated in a layer-by-layer manner
with the entire graph data: 
before start calculating the embeddings at the $\ell$-th layer,
one needs to calculate the $(\ell-1)$-layer embeddings first.
At the same time, calculating the embeddings for a layer involves performing
one-hop information propagation in the whole graph. 
Thus, based on the sequential execution paradigm,
model parallelism will suffer 
from the resource under-utilization.
For the example in Figure~\ref{fig:dist_training_model_parallel},
GPU $3$ cannot start calculating the embeddings of layer 5-6 until 
the embeddings after the first four layers are produced.
As a result, the utilization of GPU 3 is only 33\%.
In a nutshell, there is no true parallelism due to the
layer-by-layer paradigm---none of the GPUs are performing 
the computation concurrently.

\begin{figure}[htbp]
    \centering
    \includegraphics[width=0.8\linewidth]{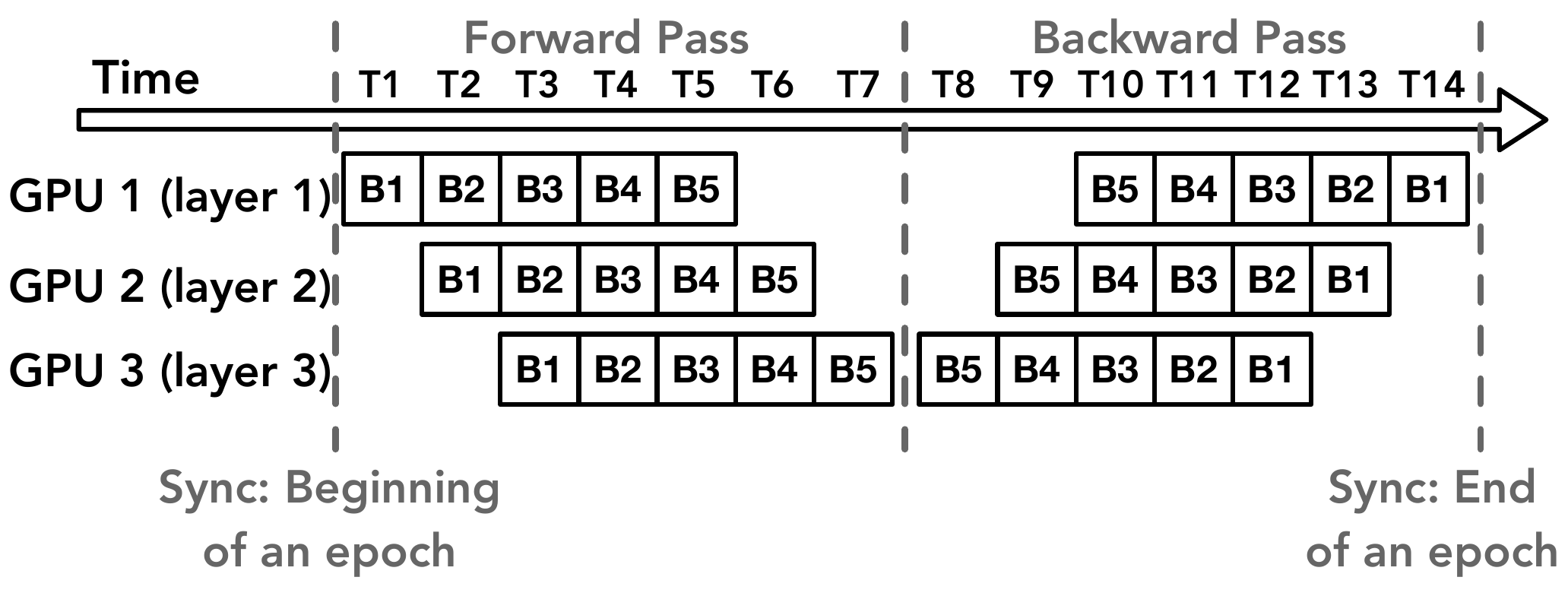}
    \vspace{-4mm}
    \caption{The Pipelining Method}
    \label{fig:gpipe}
\end{figure}

The readers may wonder why it is not an issue in recent work. 
For non-GNN neural network models, the training samples are 
{\em independent}, thanks to this property,
the resource under-utilization problem can be avoided by 
letting each GPU concurrently processing different layers 
on different training samples with delicate scheduling~\cite{huang2019gpipe,narayanan2019pipedream,narayanan2021memory,li2021chimera}. 
For example, in GPipe~\cite{huang2019gpipe},
the training samples are split into multiple ``micro-batches'' (e.g., $B1-B5$ in Figure~\ref{fig:gpipe})
and fed to the GPUs in a pipelined fashion.
As a result, after the pipeline is filled---preferably during  
most of the execution time---the GPUs can concurrently 
process different layers on different micro-batches, 
leading to a higher GPU utilization and parallelism.
For example, in Figure~\ref{fig:gpipe},
at time interval $T3$, GPU $1$, $2$, and $3$ concurrently processes micro-batches $B3$, $B2$ and $B1$, respectively.

\begin{algorithm}
\caption{The chunk-based pipelining algorithm with embedding staleness on GPU $i$.}
\label{alg:chunk_based_training}
\footnotesize
\begin{algorithmic}[1]
    \STATE \textbf{Input:} The graph $\mathcal{G}=(\mathcal{V}, \mathcal{E})$, the number of epoches $\mathcal{T}$, the layers assigned to the current GPU $\ell_i^{begin}-\ell_i^{end}$, the number of chunks $\mathcal{K}$, the learning rate $\eta$, the initial weights $\bm{W}^{(\ell_i^{begin},0)}$, $\bm{W}^{(\ell_i^{begin}+1,0)}$,
    $\bm{W}^{(\ell_i^{begin}+2,0)}$, $\ldots$,
    $\bm{W}^{(\ell_i^{end},0)}$.
    \STATE \textbf{Output:} The trained model parameters $\bm{W}^{(\ell_i^{begin},\mathcal{T})}$, $\bm{W}^{(\ell_i^{begin}+1,\mathcal{T})}$,
    $\bm{W}^{(\ell_i^{begin}+2,\mathcal{T})}$, $\ldots$,
    $\bm{W}^{(\ell_i^{end},\mathcal{T})}$
    \STATE partition $\mathcal{V}$ into $\mathcal{K}$ chunks, denoted as $\mathcal{C}_1$, $\mathcal{C}_2$, $\ldots$, $\mathcal{C}_{\mathcal{K}}$
    \FOR{$t\gets 1...\mathcal{T}$}
        \STATE \textbf{SyncAllGPUs}() // wait until all other GPUs enter the same epoch
        \STATE // start the forward pass pipeline
        \STATE $\mathcal{V}_{processed}\gets \emptyset$ 
        \FOR{$k\gets 1...\mathcal{K}$}
            \STATE $\mathcal{V}_{processed}\gets \mathcal{V}_{processed}\cup \mathcal{C}_k$
            \IF{$i>1$} 
                \STATE receive all embeddings $\bm{h}_v^{(\ell_i^{begin}-1,t)}$ ($v\in \mathcal{C}_k$) from GPU $i-1$
            \ENDIF
            \FOR{$\ell \gets \ell_i^{begin}...\ell_i^{end}$}
                \FOR{$v\in \mathcal{C}_k$}
                    \STATE $\bm{z}_v^{(\ell,t)}\gets \textsc{AGGREGATE}^{(\ell)}(\{\bm{h}_u^{(\ell-1,t)}|u\in\mathcal{N}(v)\cap\mathcal{V}_{processed} \}\cup \{\bm{h}_u^{(\ell-1,t-1)}|u\in\mathcal{N}(v)-\mathcal{V}_{processed} \})$
                    \STATE $\bm{h}_v^{(\ell,t)}\gets \textsc{UPDATE}^{(\ell)}(\bm{h}_v^{(\ell-1,t)}, \bm{z}_v^{(\ell,t)})$
                \ENDFOR
            \ENDFOR
            \IF{$i<\mathcal{K}$}
                \STATE send all embeddings $\bm{h}_v^{(\ell_i^{end},t)}$ ($v\in \mathcal{C}_k$) to GPU $i+1$
            \ENDIF
        \ENDFOR
        \STATE // start the backward pass pipeline to calculate gradients $\nabla\bm{W}^{(\ell_i^{begin},t-1)}$, $\ldots$,
        $\nabla\bm{W}^{(\ell_i^{end},t-1)}$; similiar to the forward pass; omitted for simplicity
        \STATE $\bm{W}^{(\ell,t)}\gets \bm{W}^{(\ell,t-1)} - \eta\nabla\bm{W}^{(\ell,t-1)}$ for $\ell\in [\ell_i^{begin},\ell_i^{end}]$ // update the model weights
    \ENDFOR
\end{algorithmic}
\end{algorithm}

This challenge is unique for GNNs because 
the training data is inherently {\em dependent}---the vertices are
connected to each other in the graph. 
Assume we take the similar approach as GPipe and 
split the graph vertices into 5 micro-batches
and schedule them 
as in Figure~\ref{fig:gpipe}, 
a vertex in $B1$ may have some neighbors in different micro-batches
among $B3$, $B4$, and $B5$.
Thus, calculating the layer-2 embeddings of $B1$
may require some layer-1 embeddings of $B3-B5$ of the current epoch.
However, when GPU $2$ schedules the calculation of $B1$'s 
layer-2 embeddings at time interval $T2$,
the layer-1 embeddings of $B3-B5$ have not been produced yet.
In the following two sections, we present our solution to 
achieve parallelism and training efficiency at the same time.

\subsection{Graph Chunks with Embedding Staleness}
\label{sec:chunks}






To tackle the challenge,  
we propose to slightly trade the training (statistical) efficiency for
high execution efficiency
by allowing the use of {\em stale} historical embeddings---embeddings from a previous epoch~\cite{chen2017stochastic,fey2021gnnautoscale}---for pipelined model parallelism.
During the execution,
if the calculation of an embedding $\bm{h}_v^{(\ell)}$ ($\ell$ is the GNN layer)
requires some embeddings $\bm{h}_u^{(\ell-1)}$ ($u\in\mathcal{N}(v)$) ($u$ is a 
neighbor of $v$) that have
not been produced,
we use $\bm{h}_u^{(\ell-1)}$'s value from the previous epoch
to calculate $\bm{h}_v^{(\ell)}$,
rather than stalling the pipeline to wait for the value of
$\bm{h}_u^{(\ell-1)}$ to be produced in the current epoch.

By introducing stale historical embeddings in 
pipelined execution
we derive a new GNN distributed training algorithm 
shown in Algorithm~\ref{alg:chunk_based_training} based on 
layer-level model parallelism.
For clarity,
we denote the layer-$\ell$ embedding of vertex $v$ produced in the $t$-th epoch as $\bm{h}_v^{(\ell,t)}$; and
the $\ell$-layer model weights at the end of epoch $t$ as $\bm{W}^{(\ell,t)}$.
At the beginning (line 3),
we partition the vertex set $\mathcal{V}$ into $\mathcal{K}$ chunks, which
are similar to the micro-batches in GPipe.
We use a locality-aware partitioner based on METIS~\cite{karypis1997metis} for chunk partitioning, and have $\mathcal{K} = 4\mathcal{M}$.
In each epoch, among all GPUs,
we schedule the execution of the $\mathcal{K}$ chunks
in a pipelined manner.
Once a GPU $i$ (except for the last GPU) finishes processing a chunk $\mathcal{C}_k$ for its assigned layers ($\ell_i^{begin}-\ell_i^{end}$),
it immediately sends the boundary embeddings of this chunk to the next GPU $i+1$ (line 19-21). 
Once GPU $i+1$ receives (line 10-12) the embeddings of $\mathcal{C}_k$,
it can start processing $\mathcal{C}_k$ (line 13-18) while GPU $i$ continues to process $\mathcal{C}_{k+1}$ concurrently.
In this way, all GPUs can be fully utilized when the pipelined is filled.
We track the processed vertices in a variable $\mathcal{V}_{processed}$ (line 9) so that 
we can determine whether to use historical embeddings.
For the calculation of the $\textsc{AGGREGATE}$ function (line 15),
if the current-epoch value of a neighbor-vertex embedding $\bm{h}_u^{(\ell-1,t)}$
has not been produced, i.e., $u\notin \mathcal{V}_{processed}$,
we will use its historical version $\bm{h}_u^{(\ell-1,t-1)}$ instead as an approximation.



\subsection{Training Techniques}
\label{sec:tricks}


The staleness of historical embeddings 
may slow down the convergence speed
or negatively affect the final model accuracy.
To mitigate the negative effects of the staleness of the historical 
embeddings, we propose three training techniques. 

The first technique is {\em chunk shuffling}, which randomly
shuffle the processing order of the chunks at the beginning of each epoch.
By shuffling, we can avoid the {\em systematical asymmetry} across
the chunks, e.g., $\mathcal{C}_1$ always 
suffers from more staleness compared to $\mathcal{C}_{\mathcal{K}}$.
As a result, each chunk will get more or less the same staleness during
training. 

The second technique is {\em fixing historical embeddings} to 
improve training consistency. 
Specifically, in the current epoch, 
we do not use the most recent previous epoch's 
historical embeddings, i.e., $\bm{h}_v^{(\ell-1,t-1)}$. 
Instead, we want to let the chunks in a {\em range of epoches} to use 
the {\em same} stale version of historical embeddings. 
We can express the technique concisely by introducing a 
a multiplier of $\alpha$.
In epoch $t$,
we use the historical embeddings from epoch $\alpha\lfloor(t-1)/\alpha\rfloor$.
For example, if $\alpha=10$,
during epoch $11-20$,
we always use the historical embeddings from epoch $10$.
Although doing that would seemingly 
increase the staleness since we used historical embeddings that are
not the most recent,
in practice, it improves both the convergence speed
and the final model accuracy because it allows the chunks to use
more consistent historical embeddings.

The last technique is to simply {\em avoid using historical gradients}.
The gradients are accessed in the backward pass similar to how the embeddings
are accessed in the forward pass. 
In the forward pass, the embeddings are propagated from a vertex $v$'s neighbor $u$ to $v$; while in the backward pass, the gradients are back propagated from 
$v$ to $u$. 
Despite the similarity, we observe that the historical gradients 
incurred a much larger error than historical embeddings.
\red{
Thus,
we simply omit the historical embedding gradients
in the backward pass.
When calculating the gradients of a vertex $u$,
if its neighbor vertex $v$ has not been processed yet due to the pipelined execution,
we simply replace the gradients flowing from $v$ back to $u$ with zeros.
}

\subsection{Hybrid Parallelism}
\label{sec:hybrid_analysis}

The cautious readers likely have the lingering concern that with the 
layer-level model parallelism, while the communication volume is reduced, 
it cannot support the large graph when the memory of a single GPU cannot 
accommodate the whole graph. 
The graph parallelism can naturally handle this scenario. 
We address this real issue by combining the pipelined model parallelism 
with graph parallelism, providing a general {\em hybrid} training
approaching that can reduce to layer-level model parallelism as a special case.

\begin{figure}[htbp]
    \centering
    \includegraphics[width=\linewidth]{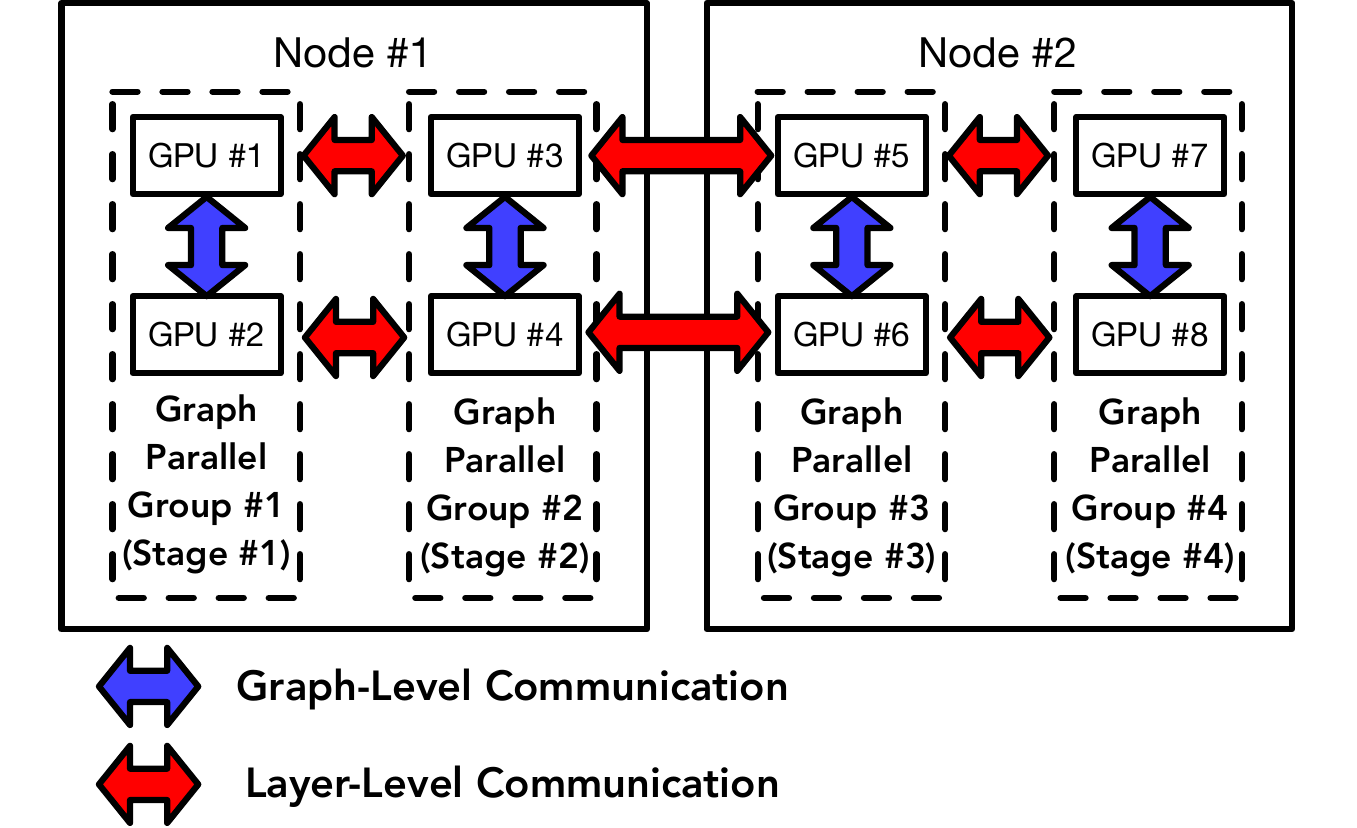}
    \vspace{-4mm}
    \caption{Hybrid parallelism between graph parallel and layer-level 
    model parallelism.}
    \vspace{-6mm}
    \label{fig:hybrid_parallel}
\end{figure}

The hybrid approach can be supported with the {\em grouping} mechanism illustrated in Figure~\ref{fig:hybrid_parallel}.
The GPUs are grouped into multiple 
size-$\mathcal{G}$ groups with graph parallelism used within each group.
To combine with pipeline parallelism, 
the number of such groups is the same as the number of desired pipeline stages.
In this setting, each group handles a single pipeline stage while the GPUs 
within the same group will process each graph partition with graph parallelism.
A given GPU still processes the graph at chunk granularity as described
in \red{Section~\ref{sec:chunks}}.
Specifically, we can assign the rank to the GPUs in each group, for the 
GPUs with the {\em same rank} in all groups, they hold the {\em same} graph 
partition, and among these GPUs, they essentially form a pure layer-level
model parallelism pipeline to process that graph partition. 
Thus, if each group just contains one GPU, the hybrid parallelism 
degenerates into the layer-level model parallelism.
Considering the graph chunks,
when a chunk $\mathcal{C}_k$ is fed to a pipline stage,
the $\mathcal{G}$ GPUs within the corresponding group first partition 
the chunk into $\mathcal{G}$ sub-chunks
(\red{the partitioning can be done as a preprocessing step to reduce cost}),
and each GPU processes a sub-chunk in parallel.
Because the GPUs in the same group leverage graph parallelism,
they will 
exchange the boundary vertex embeddings for 
each layer and hence incur additional graph communication.

In fact, the hybrid parallelism is useful for two other reasons. 
First, it is needed 
when we have more GPUs than the number of layers in a shallow GNN---a good
GNN training system should work efficiently for both deep and shallow GNNs.
Second, it is also needed for the ultra-deep GNNs for two reasons, 
either we have less number of GPUs than the number of GNN layers, or, in a
more subtle case, even with the same number of GPUs as the GNN layers, 
we may want to use hybrid parallelism if the deep pipeline still affects 
convergence after applying the aforementioned training techniques.

The hybrid parallelism allows exploiting the trade-offs
of the two types of communication.
We analyze the actual amount of both communication and show 
that it depends on both various factors. 
For the very sparse graphs, the communication volume is far from the 
worse-case complexity shown before. 
To perform the analysis, we introduce 
{\em replication factor} $\alpha$, indicating the average number of 
replicas for a vertex, which captures the number of vertices in the partition
boundaries. The replication factor is related to how the graph is partitioned
and the number of partitions.

For hybrid parallelism, assume we have $\mathcal{S}$ pipeline stages ($\mathcal{S}>1$), 
all stages have $\mathcal{W}$ graph partitions, the graph has $\mathcal{N}$ vertices, the 
GNN has $\mathcal{H}$ hidden units, and $\mathcal{L}$ layers.
For a system with $\mathcal{M}$ GPUs devoted to GNN training, we assume $\mathcal{M}=\mathcal{W}\mathcal{S}$.
We denote the replication factor in hybrid, graph, and
layer-level model parallelism as $\alpha_{h}$, $\alpha_{g}$, and 
$\alpha_{p}$, respectively.
We also denote the number of pipeline stages in hybrid and layer-level
model parallelism as $S_g$ and $S_p$, respectively.
For hybrid parallelism, the total cross-GPU communication volume is
$2\alpha_{h}\mathcal{L}\mathcal{N}\mathcal{H}+2(\mathcal{S}_h-1)\mathcal{N}\mathcal{H}$. The first term captures graph communication, $\alpha$ is determined by graph
partition and $\mathcal{W}$, larger $W$ leads to larger $\alpha$.
The second term indicates the inter-layer communication. 
The coefficient $2$ counts both forward and backward pass.
The communication for graph and layer-level model parallelism is
$2\alpha_g\mathcal{L}\mathcal{N}\mathcal{H}$ and 
$2(\mathcal{S}_p-1)\mathcal{N}\mathcal{H}$, respectively.
We can see that they are simply the individual term of the communication 
volume for hybrid parallelism, indicating that they are both special cases.

We can see that, if $\alpha_g\mathcal{L} < (\mathcal{S}_p-1)$, graph parallelism
is better than layer-level model parallelism, 
it can happen when the graph is very sparse ($\alpha_g$ is very small).
The relation between hybrid parallelism and the other two is more subtle, because
the change of setting will affect $\alpha$.
Based on the calculation, 
if $\alpha_{h}\mathcal{L}+(\mathcal{S}_h-1)<\alpha_{g}\mathcal{L}$, then 
hybrid is better than graph parallelism. 
It is possible since the larger number partition in graph parallelism may 
lead to a larger $\alpha_{g}$, making the right-hand side larger even if 
it just has one term. 
Similarly, if $\alpha_{h}\mathcal{L}+(\mathcal{S}_h-1)<(\mathcal{S}_p-1)$,
hybrid is better than layer-level model parallelism. 
It is possible because $\mathcal{S}_p$ is larger than $\mathcal{S}_g$, with
certain $\alpha_{g}$, it is possible that the right-hand side becomes larger.
Based on the above analysis, theoretically, the hybrid parallelism 
may incur the least communication. 

In our experimental results, we indeed see for certain very sparse graphs,
graph parallelism is the best, otherwise, layer-level model parallelism 
wins. We did not encounter a case where hybrid is the best, as we see from 
the above, it is based on the complex interactions among multiple factors. 
It is certainly {\em not} 
to say that the hybrid is useless, because with new graph
data sets with different characteristics and different graph partitioner, 
such case is indeed possible. 
Nevertheless, it is important that 
the system can efficiently support all the three settings for two reasons:
(1) when all settings are possible, it allows the trade-offs to be exploited; and
more importantly, (2) when the hybrid parallelism is {\em required} 
(e.g., graph is too large, more GPUs than layers, or cannot
have too deep pipeline), the system can indeed support such execution.


\section{Implementation}

\begin{figure}
    \centering
    \includegraphics[width=0.8\linewidth]{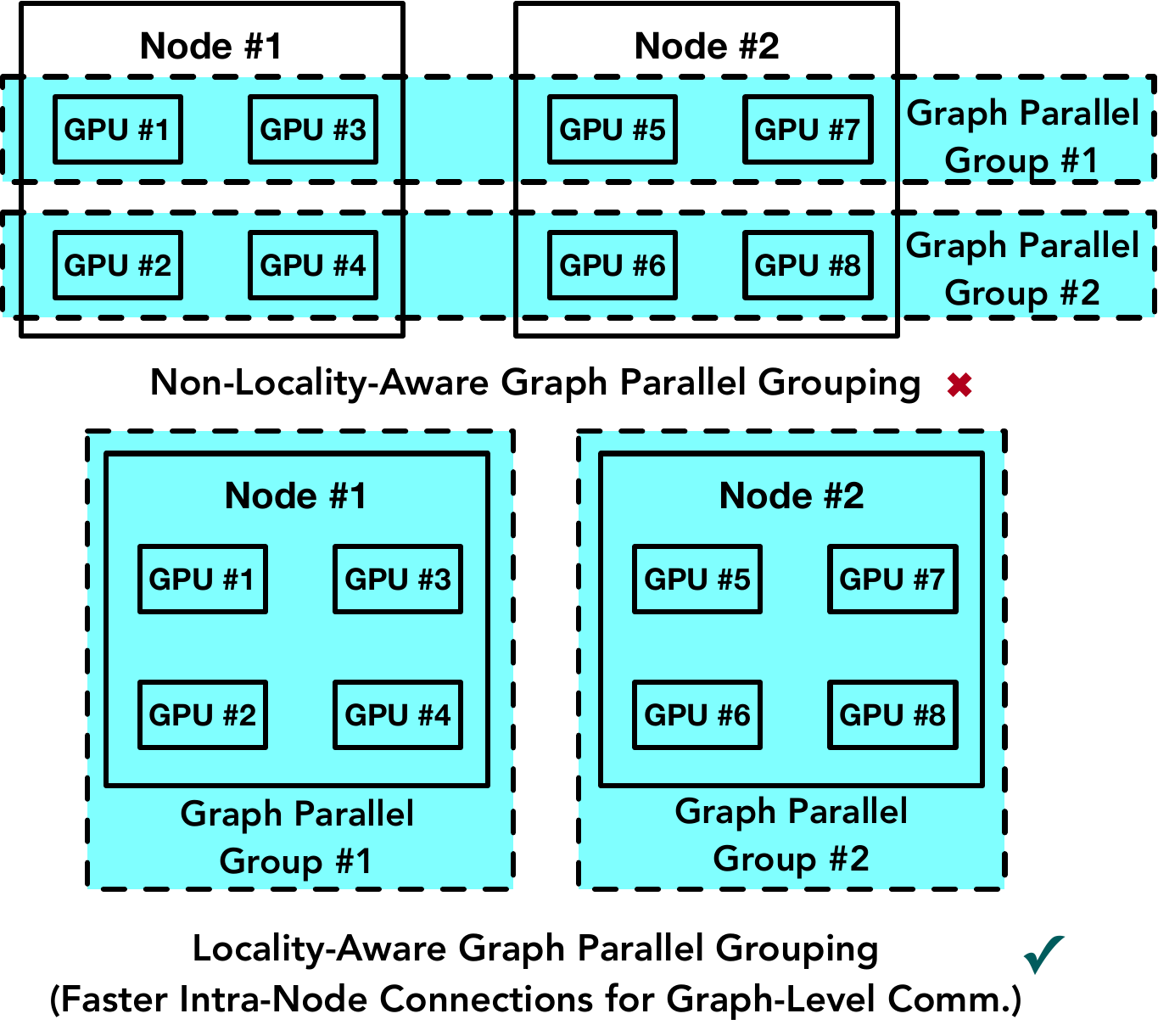}
    \vspace{-4mm}
    \caption{GPU grouping policies for hybrid parallelism.}
    \label{fig:gpu_placement}
\end{figure}

\noindent\textbf{Mapping computation to GPUs.} After the parallelism setting is determined, the system needs
to map the computation to the GPUs. 
Typically, a compute node contains multiple GPUs, and the communication
between them tends to be faster than inter-node communication, especially
with NVLink. 
For graph or layer-level model parallelism, the communication pattern between
different partitions or pipeline stages is symmetry, thus there is no 
special considerations when they are mapped to certain GPUs. 

For the hybrid parallelism, the
GPU grouping policy is performance-critical because the communication
between graph partitions is usually more intensive and irregular compared to the 
static and predictable communication pattern of inter-layer communication. 
To accommodate this observation, a key principle is to maximize the GPU locality within the same group.
For example, 
in Figure~\ref{fig:gpu_placement},
to group 8 GPUs from 2 nodes into 2 groups,
grouping choice \{\{1, 2, 3, 4\}, \{5, 6, 7, 8\}\} is more preferable than
\{\{1, 3, 5, 7\}, \{2, 4, 6, 8\}\}.
The first grouping choice assign the GPUs from the same node to the same group, which allows the intensive intra-group graph communciation
go through the much faster intra-node links like PCIe or NVLink.

\noindent\textbf{Implementation details.}
\red{\proj is implemented in C++ on top of CUDA, cuDNN and cuBLAS with roughly 16K lines of code.
We use roughly 5K lines of code 
to implement different neural network operators needed by GNNs (e.g., Layer Normalization~\cite{ba2016layer})
and SGD-based model optimizers like Adam Optimizer~\cite{kingma2014adam},
roughly 2K lines of code to
implement graph data management,
and 8K lines of code to 
implement the scheduler
and the communication subsystem
that support three types of parallelism.
We use NCCL~\cite{jeaugey2017nccl} for inter-GPU communication
and use MPI for process management.}



%% file: experiments.tex
\section{Experiments}

\subsection{Experiment Settings}

By default, the experiments are evaluated on a testbed with two GPU nodes,
each of which contains four NVIDIA A5000 GPUs, two AMD EPYC 7302 16-core CPUs and 256GB DDR4 RAM.
The GPUs within the same node are connected with PCIe 4.0x16 while  
a 200Gbps InfiniBand network
is used for cross-node 
communication.

We use four datasets shown in Table~\ref{tab:dataset}, including
Squirrel~\cite{rozemberczki2021multi}, 
Physics~\cite{shchur2018pitfalls},
Flickr~\cite{zeng2019graphsaint} and Reddit~\cite{hamilton2017inductive}.
We also show the replication factors of the 
datasets (8 partitions with METIS~\cite{karypis1997metis}).
Squirrel, Flickr and Reddit are relatively densely-connected datasets---their replication factors are more than 2.
\red{In contrast, Physics is easier to partition, so its replication factor is small than 1}.
We choose datasets with various replication factors to analyze the effect of graph properties on our methods.
We evaluate \proj with four models: GCN~\cite{kipf2016semi}, GraphSage~\cite{hamilton2017inductive}, GCNII~\cite{chen2020simple},
and ResGCN+~\cite{li2021training,li2020deepergcn}.
Unless otherwise mentioned,
the model depths are set to 32 layers as our approach mostly focuses on deep GNNs.
Note that GCN and GraphSage are not originally designed to be deep, 
thus they do not converge well to a reasonable accuracy with 32 layers.
Hence, we only use these two models for performance analysis.
GCNII and ResGCN+ are GNN models designed to be deep. Hence we use them for both performance and training accuracy analysis.
The numbers of hidden units are 1000 for small graphs (Squirrel) and reduced to 100 for larger datasets (Physics, Flickr and Reddit).
We use Adam optimizer~\cite{kingma2014adam} to train the models and the learning rate is set to the default value (0.001).
The dropout rate is 0.5 and the number of training epoches is 5000.
For comparison purpose,
we also implemented a baseline with graph parallelism on top of the same software stack.
For the baseline \red{and hybrid parallelism}, the graphs are partitioned by the METIS partitioner~\cite{karypis1997metis} to minimize inter-partition communication. 
For \proj, we use two settings throughout the evaluation: 1) pure pipelined layer-level model parallelism: the model is divided into 8 pipeline stages \red{with 4 layers in each stage}, and each GPU handles one stage; 2) hybrid parallelism: the model is divided into 4 pipeline stages and each stage is handle by two GPUs. The GPUs within the same stages leverage graph parallelism.



\begin{table}[htbp]
    \centering
    \scalebox{0.73}{
    \begin{tabular}{ccccccc}
         \hline
         Dataset & \#Vertices & \#Edges & \#Features & \#Classes & $\mathcal{\alpha}$ & Avg.Degree \\
         \hline
         Squirrel & 5.2K & 396.7K & 2089 & 5 & 2.22 & 76.3 \\ 
         Physics & 34.5K & 495.9K & 8415 & 5 & 0.99 & 14.4 \\
         Flickr & 89.3K & 899.8K & 500 & 7  & 2.15 & 10.1 \\ 
         Reddit & 233.0K & 114.6M & 602 & 41 & 2.61 & 491.8 \\
         \hline
    \end{tabular}}
    \caption{Graph datasets, $\mathcal{\alpha}$ is the replication factor with 8 partitions.}
    \label{tab:dataset}
\end{table}

\subsection{Training Efficiencies}


\noindent\textbf{Comparing with graph parallelism baseline.}
We evaluate the training efficiency of \proj
by comparing its per-epoch training time with
the baseline using graph parallelism in 
Table~\ref{tab:per_epoch_time}.
On Squirrel, Flickr, and Reddit,
\proj (\red{indicated as Pipeline}) is able to significantly reduce the per-epoch training time by up to \maxspeedupgraphparallel.
The speedups are attributed to the reduction in communication volume thanks to \proj's lower communication complexity.
It is worth noting that the performance of the hybrid parallelism is also better than graph parallelism, but slightly worse than pure pipeline parallelism.
It is because of the additional graph-level communication within each size-2 graph parallel group.
However, as those additional communication only go through the fast intra-node links (PCIe 4.0x16)
rather than the slower network adapters,
the additional communication overhead is not significant.

We also note that on Physics,
\proj is slower than the baseline using graph parallelism.
It is because the computation workload on the Physics dataset
is very lightweight.
\proj adopts the chunk-based pipelining method, i.e., it cuts the lightweight workload into
multiple small chunks and executes them one by one
on GPU.
The workload of each chunk is too small (usually less than 10ms),
which hurts the GPU utilization and 
degrade the performance.

\begin{table}[htbp]
    \centering
    \scalebox{0.95}{
    \begin{tabular}{c|c|c|c|c}
        \hline
        Dataset & Model & Graph & Pipeline & Hybrid \\
        \hline
        \multirow{4}{*}{Squirrel} & GCN & 0.19 & 0.08 & 0.11 \\
        & GraphSage & 0.27 & 0.15 & 0.19 \\
        & GCNII & 0.20 & 0.11 & 0.14 \\
        & ResGCN+ & 0.27 & 0.19 & 0.24 \\
        \hline
        \multirow{4}{*}{Physics} & GCN & 0.05 & 0.07 & 0.08 \\
        & GraphSage & 0.07 & 0.13 & 0.12 \\
        & GCNII & 0.06 & 0.11 & 0.10 \\
        & ResGCN+ & 0.08 & 0.14 & 0.15 \\
        \hline
        \multirow{4}{*}{Flickr} & GCN & 0.16 & 0.06 & 0.09 \\
        & GraphSage & 0.17 & 0.10 & 0.13 \\
        & GCNII & 0.17 & 0.09 & 0.12 \\
        & ResGCN+ & 0.19 & 0.16 & 0.19 \\
        \hline
        \multirow{4}{*}{Reddit} & GCN & 0.87 & 0.36 & 0.39 \\
        & GraphSage & 1.08 & 0.53 & 0.57 \\
        & GCNII & 0.90 & 0.41 & 0.45 \\
        & ResGCN+ & 0.93 & 0.51 & 0.55 \\
        \hline
    \end{tabular}}
    \caption{Comparing the per-epoch training time (unit: s) between the baseline using graph parallelism and \proj (Graph: graph parallelism; Pipeline: layer-level model parallelism with 8 pipeline stages; Hybrid: hybrid parallelism of graph parallel and pipeline parallel with 2 graph parallel ways and 4 pipeline stages).}
    \label{tab:per_epoch_time}
\end{table}

\noindent \textbf{Comparing with DGL.}
We also compares \proj (pipelined layer-level model parallel) with DGL~\cite{wang2019deep}, a state-of-the-art system supporting mini-batch-based distributed training.
We choose two datasets (Squirrel and Flickr) and two models (GCN and GraphSage) for the comparison.
Since the evaluated datasets can entirely 
fit into the GPU memory,
we configure DGL so that each
GPU has a complete copy of the graph data
to eliminate the communication overhead caused by unnecessary graph partitioning.
We use the full neighbor sampler to construct the mini-batch so that all edges will be used for each epoch.
The batch size is set to 256.
We show the results in Table~\ref{tab:comp_dgl}.
\proj is able to significantly outperform DGL
\red{by one order of magnitude}.
Since we disable graph partitioning for DGL,
its performance inferiority
is mostly due to the
computation redundancy:
as discussed earlier in the paper,
for deep models,
each mini-batch largely overlaps with 
each other,
and hence the computation due to such overlap 
is performed multiple times.

\begin{table}[htbp]
    \centering
    \scalebox{1.}{
    \begin{tabular}{c|c|c|c|c}
         \hline
         Dataset & Model & DGL & \proj & Speedup \\
         \hline
         \multirow{2}{*}{Squirrel} & GCN & 1.52 & 0.08 & 19.0 \\
         & GraphSage & 1.57 & 0.15 & 10.5 \\
         \hline
         \multirow{2}{*}{Flickr} & GCN & 3.66 & 0.06 & 61.0 \\
         & GraphSage & 3.98 & 0.10 & 39.8 \\
         \hline
    \end{tabular}}
    \caption{Comparing with DGL (unit: s).}
    \label{tab:comp_dgl}
\end{table}


\noindent\textbf{Communication analysis.}
We further confirm the communication superiority of \proj by the communication 
analysis shown in Table~\ref{tab:comm_volume} and Table~\ref{tab:comm_overhead}.
\proj with pure pipelined layer-level model 
parallelism is able to significantly
reduce the communication volume and communication overhead by up to 
\maxcommvolumereduction and \maxcommoverheadreduction (on average \avgcommvolumereduction and \avgcommoverheadreduction), respectively.

\begin{table}[htbp]
    \centering
    \scalebox{0.95}{
    \begin{tabular}{c|c|c|c|c}
        \hline
        Dataset & Model & Graph & Pipeline & Hybrid \\
        \hline
        \multirow{4}{*}{Squirrel} & GCN & 4.43 & 0.27 & 1.38 \\
        & GraphSage & 6.10 & 0.27 & 1.61 \\
        & GCNII & 4.53 & 0.54 & 1.51 \\
        & ResGCN+ & 6.20 & 0.27 & 1.63 \\
        \hline
        \multirow{4}{*}{Physics} & GCN & 0.88 & 0.18 & 0.62 \\
        & GraphSage & 0.94 & 0.18 & 0.63 \\
        & GCNII & 0.88 & 0.36 & 0.70 \\
        & ResGCN+ & 0.89 & 0.18 & 0.62 \\
        \hline
        \multirow{4}{*}{Flickr} & GCN & 4.60 & 0.47 & 2.00 \\
        & GraphSage & 4.62 & 0.47 & 2.00 \\
        & GCNII & 4.60 & 0.93 & 2.19 \\
        & ResGCN+ & 4.62 & 0.47 & 2.00 \\
        \hline
        \multirow{4}{*}{Reddit} & GCN & 14.50 & 1.22 & 5.87 \\
        & GraphSage & 14.52 & 1.22 & 5.87 \\
        & GCNII & 14.50 & 2.43 & 6.39 \\
        & ResGCN+ & 14.52 & 1.22 & 5.87 \\
        \hline
    \end{tabular}}
    \caption{Comparing the per-epoch communication volume (unit: GB) between the baseline using graph parallelism and \proj.}
    \label{tab:comm_volume}
\end{table}

It is worth noting that sometimes the reduction in communication time is even more 
significant than the reduction in communication volume.
For example, for GCN on Reddit,
\proj with pipeline parallelism reduces the communication overhead by 24.7x
while the communication volume is only reduced by 11.9x.
This indicates that the communication pattern 
of the pipelined layer-level model parallelism 
is more efficient and simpler than graph parallelism:
each GPU only needs to send the data to one other GPU.
By comparison,
for graph parallelism, 
one GPU usually needs to send data to all other GPUs concurrently,
which can easily over-utilize some slow bottleneck links (e.g., inter-node links)
and hence hurt the overall communication efficiency.



\subsection{Scalability}

\begin{figure}
    \centering
    \includegraphics[width=\linewidth]{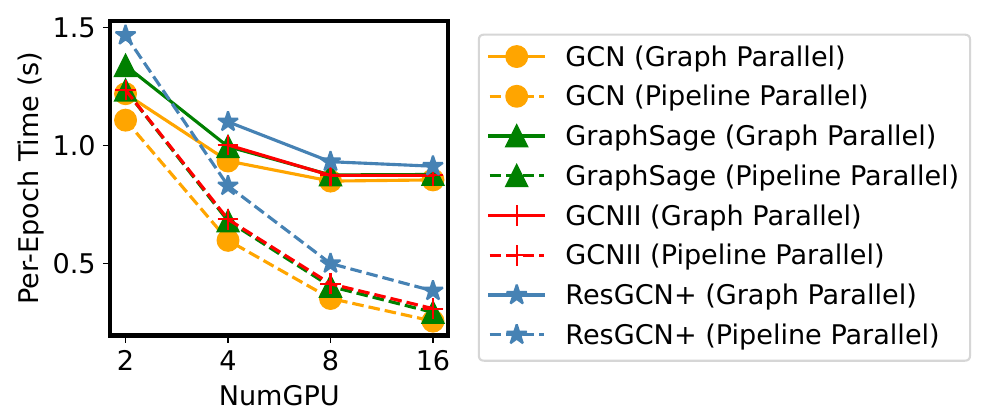}
    \vspace{-8mm}
    \caption{Scalability Analysis.}
    \label{fig:scalability}
\end{figure}

We analyze the scalability of \proj
by using various numbers of GPUs to 
train the 32-layer models
on Reddit,
and show the results in Figure~\ref{fig:scalability}.
The missing data points (e.g., the 2-GPU result of ResGCN+ with graph parallelism)
are due to out-of-GPU-memory errors.
For all evaluated models,
\proj with pipelined layer-level model parallelism
exhibits better scalability
than the graph-parallel baseline.
For example,
for GCN,
with graph parallelism,
using 16 GPUs (0.85s) is only 1.43x
faster
than using 2 GPUs (1.22s).
In contrast,
with \proj,
scaling from 2 GPUs (1.11s) to 16 GPUs (0.26s)
leads to a speedup of 4.3x.

\begin{table}[htbp]
    \centering
    \scalebox{0.95}{
    \begin{tabular}{c|c|c|c|c}
        \hline
        Dataset & Model & Graph & Pipeline & Hybrid \\
        \hline
        \multirow{4}{*}{Squirrel} & GCN & 113.75 & 7.54 & 26.03 \\
        & GraphSage & 123.11 & 7.82 & 31.05 \\
        & GCNII & 113.36 & 12.02 & 30.78 \\
        & ResGCN+ & 115.16 & 8.00 & 32.14 \\
        \hline
        \multirow{4}{*}{Flickr} & GCN & 124.32 & 8.13 & 33.38 \\
        & GraphSage & 123.52 & 9.60 & 33.69 \\
        & GCNII & 123.90 & 14.79 & 38.39 \\
        & ResGCN+ & 124.38 & 11.11 & 35.73 \\
        \hline
        \multirow{4}{*}{Reddit} & GCN & 634.01 & 25.66 & 78.86 \\
        & GraphSage & 714.09 & 26.24 & 82.21 \\
        & GCNII & 636.00 & 42.12 & 96.77 \\
        & ResGCN+ & 637.51 & 26.41 & 85.53 \\
        \hline
        \multirow{4}{*}{Physics} & GCN & 36.04 & 5.96 & 21.47 \\
        & GraphSage & 36.93 & 6.13 & 23.58 \\
        & GCNII & 36.47 & 10.95 & 25.98 \\
        & ResGCN+ & 37.40 & 6.19 & 24.63 \\
        \hline
    \end{tabular}}
    \caption{Comparing the per-epoch communication overhead (unit: ms) between the baseline using graph parallelism and \proj.}
    \label{tab:comm_overhead}
\end{table}

\subsection{Execution Time Breakdown}

\begin{figure}
    \centering
    \includegraphics[width=\linewidth]{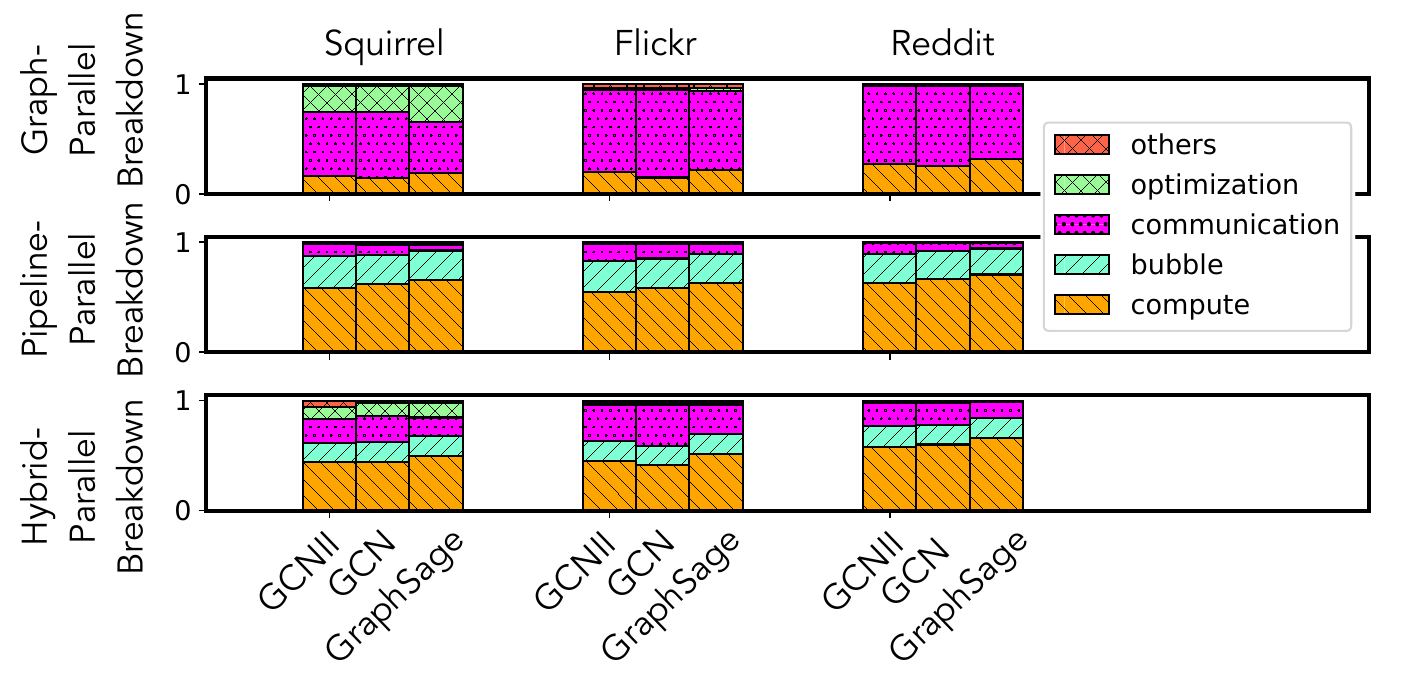}
    \caption{Training time breakdown analysis.}
    \label{fig:breakdown_analysis}
\end{figure}

We also present the execution time breakdown of \proj and the baseline in Figure~\ref{fig:breakdown_analysis}.
We use three datasets (Squirrel, Flickr, and Reddit) and three models (GCNII, GCN, and GraphSage) for the analysis.
Compared to the baseline using graph parallelism, 
whose training time is dominated by the communication overhead (on average 66.5\%),
\proj only spends 9.5\% on average on communication while its training time is dominated 
by actual computation (on average 62.5\%),
indicating a better resource utilization.
We also note that the pipelining introduces a bubble overhead of 26.5\% (on average).
The bubble overhead may be reduced by adopting a more advanced pipelining technique~\cite{li2021chimera,narayanan2019pipedream} other than GPipe. 
We leave reducing the bubble overhead as our future work.

\subsection{Sensitivity Study on Model Depth}

We also conducted sensitivity tests on the model depth
to see how it affects the 
communication volume of 
\proj (layer-level model parallelism) and the baseline using graph
parallelism.
The results are shown in Table~\ref{tab:various_depth}.
The communication of the baseline
increase almost linearly 
with the model depth
while that of \proj stays unchanged.
For example, 
on the Physics dataset,
when the model depth is 8,
the communication volume of the 
baseline is less than \proj.
However, 
as the model depth increases to 128,
the communication volume increases 
to 3.38GB,
which is 9.4x larger 
than our system.
This observation is consistent with
our analysis in 
Section~\ref{sec:pipeline_parallelism}:
the communication complexity of layer-level model parallelism is 
independent of the model depth.

\begin{table}[htbp]
    \centering
    \scalebox{0.8}{
    \begin{tabular}{c|c|c|c|c|c|c}
         \hline
         Dataset & Model Depth & 8 & 16 & 32 & 64 & 128\\
         \hline
         \multirow{2}{*}{Squirrel} & Graph Comm. (GB) & 1.22 & 2.32 & 4.53 & 8.95 & 17.80 \\
         \cline{2-7}
         & Pipeline Comm. (GB) & \multicolumn{5}{c}{0.54} \\
         \hline
         \multirow{2}{*}{Physics} &  Graph Comm. (GB) & 0.25 & 0.46 & 0.88 & 1.71 & 3.38 \\
         \cline{2-7}
         & Pipeline Comm. (GB) & \multicolumn{5}{c}{0.36} \\
         \hline
    \end{tabular}}
    \caption{\proj's per-epoch communication volume with various model depth (model: GCNII).}
    \label{tab:various_depth}
\end{table}

\subsection{GNN Convergence and Accuracy Analysis}

\begin{figure*}[htbp]
    \centering
    \includegraphics[width=\linewidth]{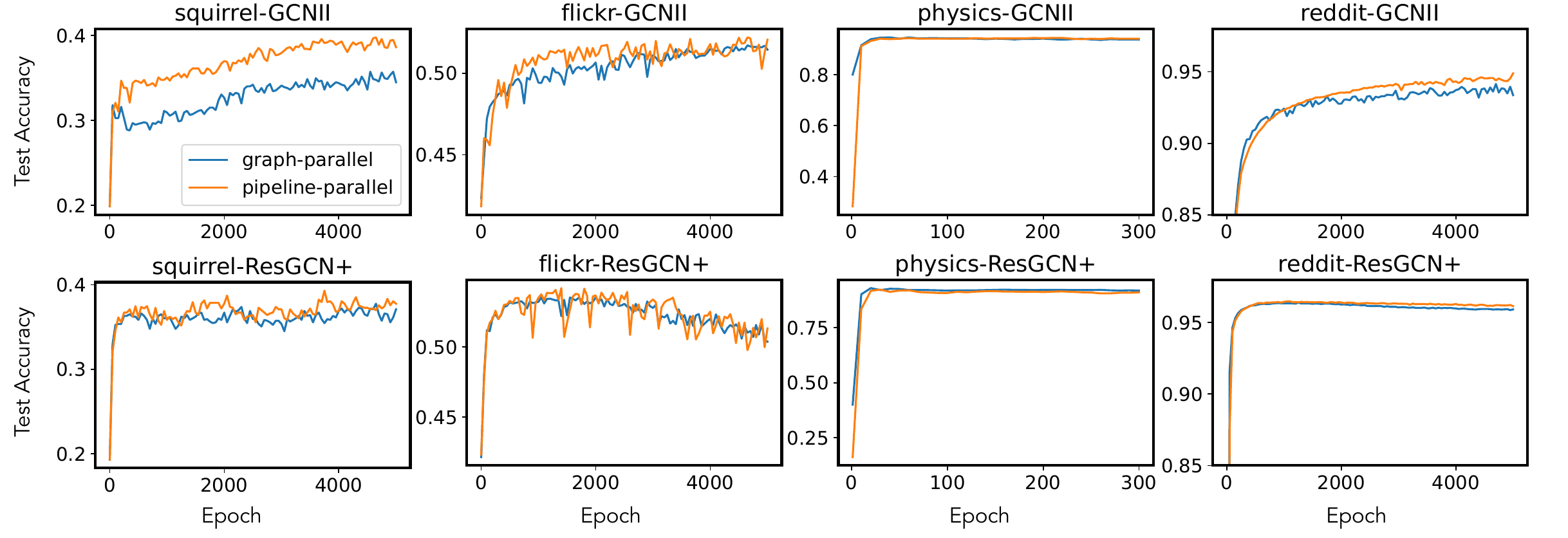}
    \vspace{-4mm}
    \caption{Compared to the baseline using graph parallelism, \proj with layer-level model parallelism can converge at a similar or faster speed and achieve a comparable (sometimes even higher) level of model accuracy.}
    \label{fig:convergence_analysis}
\end{figure*}

\begin{figure}[htbp]
    \centering
    \includegraphics[width=\linewidth]{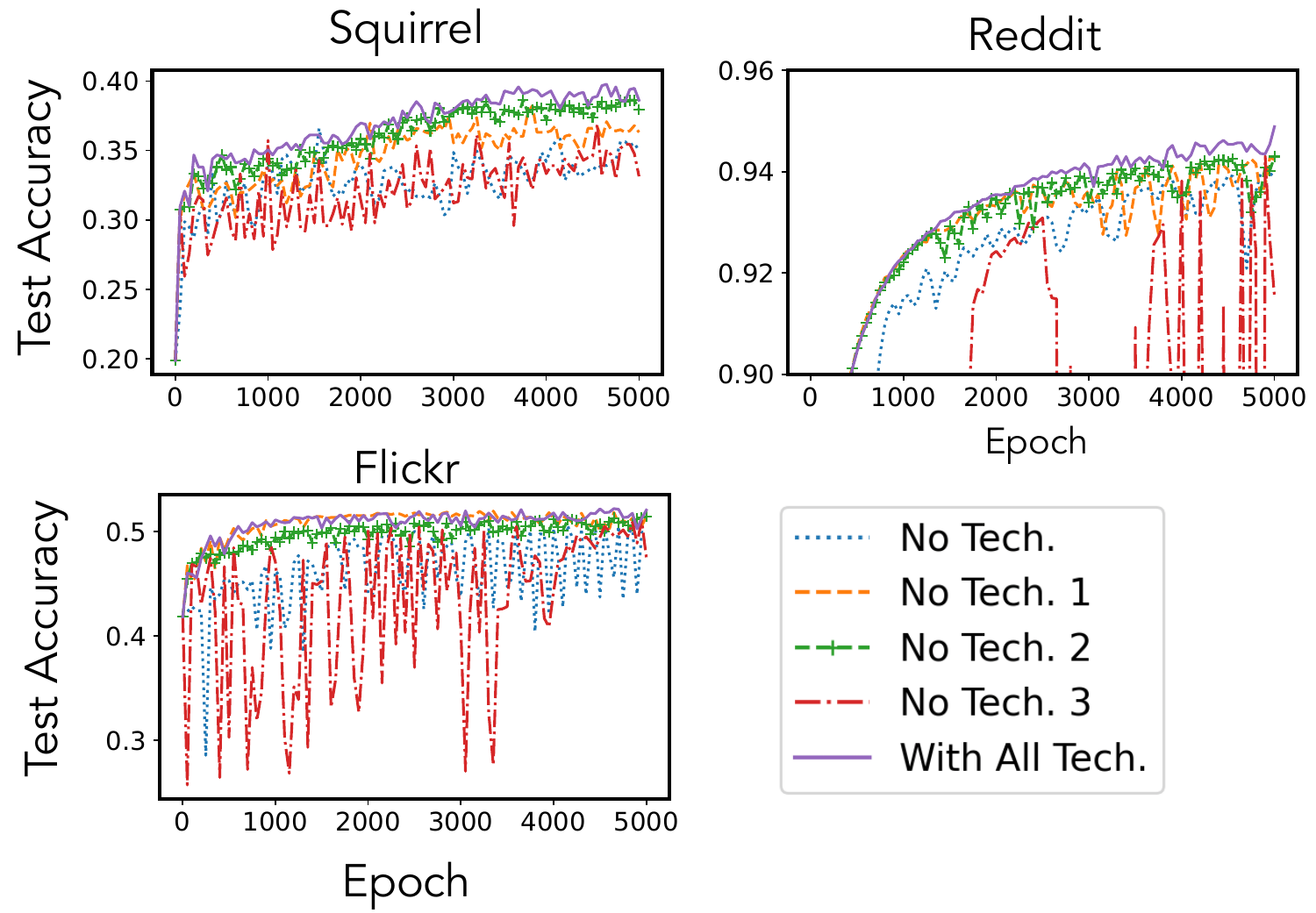}
    \vspace{-3mm}
    \caption{Effect of training techniques (GCNII).}
    \vspace{-4mm}
    \label{fig:convergence_tricks}
\end{figure}

\noindent\textbf{Comparing with graph parallelism.}
We analyze the model convergence speed and the model accuracy in Figure~\ref{fig:convergence_analysis}
for GCNII and ResGCN+ on all datasets.
We omit GCN and GraphSage since they usually cannot converge to a reasonable accuracy with 32 layers,
as they are not designed to be deep at the algorithm level.
We note that in spite of the staleness introduced 
among chunks in the pipeline,  
the model can converge 
with a similar number of epochs to the baseline using graph parallelism (or even fewer epochs at times),
and achieve a comparable level of final model accuracy
across all dataset-model combinations.
This indicates that, in practice, 
\proj does not hurt the model convergence while achieving a better training efficiency.
\noindent\textbf{Analyzing the training techniques.}
We further analyze the three training techniques in Section~\ref{sec:tricks} that aim to improve training
stability.
We compare the training curve 
of \proj with 
four of its variants 
with one or all training techniques disabled.
We perform the analysis
on Squirrel, Flickr and Reddit with 
the GCNII model
and the results are presented in Figure~\ref{fig:convergence_tricks}.
The version with all training techniques always outperforms all other variants.
It converges at the fastest speed,
exhibits the best training stability (i.e., less fluctuation),
and achieves the highest final model accuracy.
We find out that the third training technique (avoiding using historical gradients)
is the most helpful one.
In practice, 
we observe that the gradients of vertex
embeddings vary
significantly across epochs.
Hence, using the historical gradients 
in the backward pipelining incurs 
non-negligible staleness,
which causes significant accuracy drop periodically
and prevents the model from convergence.

\begin{table}[htbp]
    \centering
    \scalebox{0.8}{
    \begin{tabular}{c|c|c|c|c}
         \hline
         Model & \multicolumn{2}{c|}{Graph Parallel} & \multicolumn{2}{c}{Hybrid Parallel} \\
         \hline
         & Time (ms) & Comm. (GB) & Time (ms) & Comm. (GB) \\
         \hline
         GCN & 107 & 1.815 & 86 & 1.530 \\
         GraphSage & 112 & 1.819 & 94 & 1.532 \\
         GCNII & 112 & 1.816 & 99 & 1.704 \\
         ResGCN+ & 117 & 1.818 & 104 & 1.532 \\ 
         \hline
    \end{tabular}}
    \caption{Comparing the per-epoch training time (unit: ms) and communication volume (unit: GB) of hybrid parallelism (with 2 pipeline stages) with graph parallelism on shallow models (4-layer models, dataset: Reddit).}
    \label{tab:shallow_models}
\end{table}

\subsection{Hybrid Parallelism on Shallow GNNs}


We also analyze the performance of 
hybrid parallelism 
on shallow GNNs, 
for which pure pipelined layer-level model parallelism
does not apply since the number of layers
is smaller than the number of GPUs.
We choose to evaluate 4-layer models 
on the Reddit dataset,
and report the results 
in Table~\ref{tab:shallow_models}.
Hybrid parallelism outperforms graph parallelism 
for all models on Reddit by 1.17x on average
since it incurs less communication cost.
It is because although introducing extra layer-level communication,
hybrid parallelism partitions the graph
into fewer parts,
and hence reduces the replication factor and
the graph communication.
This may leads to a lower overall communication cost.

%% file: related_works.tex
\section{Related Works}

\noindent\textbf{Distributed Graph Processing.} 
Distributed graph processing systems~\cite{malewicz2010pregel,gonzalez2012powergraph,gonzalez2014graphx,zhu2016gemini,dathathri2018gluon,dathathri2019gluon,chen2019powerlyra,zhuo2021distributed} 
are designed for
traditional message-propagation graph analytics workloads like BFS and PageRank in a graph-parallel manner. 
Although these systems usually provide a flexible programming 
interface,
it is hard to implement GNN-based graph workloads on them 
since they in general lack nerual network supports, e.g., 
support to tensor operations and automatic differentiation.

\noindent\textbf{Distributed GNN Training.} 
Distributed systems and libraries tailored for GNN workloads can be roughly classified 
into two categories. 
The first category includes those leverage graph-level 
parallelism for \textit{full-graph} training.
NeuGraph~\cite{ma2019neugraph} is one of the earliest distributed GNN systems.
ROC~\cite{jia2020improving} 
attempts to improve the training scalability.
G3~\cite{wan2023scalable} and Dorylus~\cite{thorpe2021dorylus}
also divide the GNN workload into small chunks 
for training and exploits pipeline parallelism.
However, 
they {\em pipeline the communication and computation} of these chunks
to hide communication cost, which is different from the 
\red{pipelining among chunks at layer-level in our approach}. Each GPU (in G3) or graph server (in Dorylus) still focuses on {\em all layers} of one graph partition,
and hence suffers from the communication issue of the graph parallelism.
A lot of recent works also try to reduce or hide the massive graph-level communication cost~\cite{wan2022pipegcn,wan2022bns,peng2022sancus,mostafa2022sequential,wan2023adaptive,wang2023mgg}.
Compared to these systems or methods,
\proj proposes to exploit a {\em new dimension} of parallelism for GNN training with a lower worst-case communication complexity.
The second category includes distributed variants of the sampling-based method~\cite{zheng2020distdgl,fey2019fast,gandhi2021p3,zheng2022bytegnn,liu2023bgl,polisetty2023gsplit,yang2022gnnlab,sun2023legion,zhang2023ducati}:
each GPU processes small randomly sampled subgraphs concurrently.
These works mostly focus on improving the efficiency of subgraph sampling.

\noindent\textbf{Pipelined Model Parallelism.} 
GPipe~\cite{huang2019gpipe} and PipeDream~\cite{narayanan2019pipedream} 
are the first two works that apply piplining to 
improve GPU utilization
for model parallelism distributed training.
GPipe~\cite{huang2019gpipe} proposes to divide a mini-batch 
into multiple micro-batches
and execute the micro-batches in a pipelined manner.
More sophisticated pipelining methods are proposed later to reduce pipeline bubbles~\cite{narayanan2019pipedream,li2021chimera},
reduce memory consumption~\cite{narayanan2021memory},
and improves system efficiency~\cite{fan2021dapple}.
However, these existing works apply to non-GNN models like CNN, 
which do not have complicated inter-sample dependencies (i.e., inter-sample edges for GNN models). 
To the best of our knowledge, 
\proj is the {\em first} approach 
that points out the 
advantage of pipelined model parallelism
in terms of communication complexity
for distributed GNN training,
and provides an effective method using 
pipelined layer-level model parallelism training method 
that outperforms graph parallelism.

%% file: conclusion.tex
\section{Conclusion}

This paper proposes {\em \proj}, a new approach
that scales the distributed 
full-graph {\em deep} GNN training. 
Being the first to adopt layer-level 
model parallelism for GNN training, \proj 
partitions GNN layers among GPUs, each device is 
responsible for performing the computation
for a disjoint subset of consecutive GNN layers
on the whole graph. 
Compared to graph 
parallelism with each GPU handling a graph 
partition, \proj reduces the communication
volume by a factor of the number of GNN layers. 
\proj overcomes the unique challenges for 
pipelined layer-level model parallelism 
on the whole graph by partitioning it into potentially dependent chunks, allowing the
use of historical vertex embeddings, and
specific training techniques to ensure convergence. 
Extensive experiments show that our approach
significantly speedups distributed GNN training
compared to graph parallelism
while achieving a comparable level of model accuracy and convergence speed.
